\documentclass[12pt]{article}
\usepackage{amsmath,amssymb}
\usepackage{cite}
\usepackage{graphicx,color}
\usepackage{setspace}
\usepackage{nicefrac}
\usepackage[nottoc]{tocbibind} 

\usepackage[linktoc=all,hidelinks]{hyperref}

\usepackage{indentfirst}

\usepackage[lf]{Baskervaldx} 
\usepackage[bigdelims,vvarbb]{newtxmath} 
\usepackage[cal=boondoxo]{mathalfa} 

\usepackage{geometry}
\DeclareSymbolFont{usualmathcal}{OMS}{cmsy}{m}{n}
\DeclareSymbolFontAlphabet{\mathcal}{usualmathcal}

\def\cR{\mathcal{R}}
\def\cL{\mathcal{L}}

\def\tR{\tilde{R}}

\newcommand{\LL}{\text{LL}}
\newcommand{\RLL}{\text{RLL}}
\newcommand{\SP}{\text{SP}}
\newcommand{\MG}{\text{MG}}

\newcommand{\link}[1]{[\href{http://arxiv.org/abs/#1}{{\tt arXiv:#1}}]}
\newcommand{\linkth}[1]{[\href{http://arxiv.org/abs/hep-th/#1}{{\tt arXiv/hep-th:#1}}]}

\newcommand{\mail}[1]{\href{mailto:#1}{{\tt #1}}}
\DeclareMathOperator*{\hlim}{h-lim}
\newcommand*\dif{\mathop{}\!\mathrm{d}}

\usepackage{titletoc}
\dottedcontents{section}[1.2em]{}{1em}{0.5pc}
\dottedcontents{subsection}[4em]{}{2em}{1pc}

\begin{document}



\begin{center}
	\setstretch{2}
	{\LARGE \bf Three Dimensional Modified Gravities as Holographic Limits of Lancsoz-Lovelock Theories}
\end{center}

\begin{center}

{G{\"o}khan Alka\c{c}$\,{}^{a}$, Deniz Olgu Devecio\u{g}lu$\,{}^{b}$}
\\[4mm]

{\small 
{\it ${}^a$Physics Engineering Department, Faculty of Engineering,\\ Hacettepe University, 06800, Ankara, Turkey}\\[2mm]

{\it ${}^b$School of Physics, Huazhong University of Science and Technology,\\
	Wuhan, Hubei,  430074, China}\\[2mm]

e-mail: {\mail{gokhanalkac@hacettepe.edu.tr}, \mail{dodeve@gmail.com}}
}
\end{center}
\noindent\rule{\textwidth}{1pt}
{\bf Abstract} We define a ``holographic limit'' of Lancsoz-Lovelock theories  at the Lagrangian level which gives three-dimensional modified gravity theories. We also show that this limit applies to more general classes of theories in higher dimensions provided that they admit the holographic c-theorem.




\noindent\rule{\textwidth}{1pt}

Despite its enormous successes in explaining gravitational phenomena, Einstein's General Theory of Relativity (GR) is considered incomplete due to its non-renormalizable nature. When quantized with standard rules of quantum field theory that we use to describe other fundamental interactions, the theory loses its predictive power at high energies, and finding its UV completion is one of the most important - if not the most important - problems in theoretical physics. One natural attempt for a solution is to consider some modifications to the theory and/or its variants in dimensions other than four. In this context, the Lancsoz-Lovelock (LL) theories play a central role \cite{Lanczos:1938sf,Lovelock:1971yv,Lovelock:1972vz}. They form the most general class of theories with covariantly conserved field equations containing no derivatives of the metric tensor of order higher than two and they share many key properties with GR (see \cite{Padmanabhan:2013xyr} for a review), notably possessing unitary massless spin-2 excitations around any of their constant curvature vacua \cite{Sisman:2012rc}. In the action formulation, LL theories are defined through polynomials of the Riemann tensor which takes the form
\begin{equation}
\cL^{\LL}_m \propto \delta^{a_1 b_1 a_2 b_2 \ldots a_m b_m}_{c_1 d_1 c_2 d_2 \ldots c_m d_m} R^{c_1 d_1}_{a_1 b_1} R^{c_2 d_2}_{a_2 b_2} \ldots R^{c_m d_m}_{a_m b_m},
\end{equation}
where $\cL^{\LL}_m$ denotes the $m$-th order LL invariant. For any order $m$, there is a critical dimension $D=2m$ where $\cL^{\LL}_m$ is a boundary term and the nontrivial contribution to the field equations arises in ${D\geq 2m + 1}$. Due to the antisymmetrization in the generalized Kronecker delta symbol, all the $\cL^{\LL}_m$'s vanish in three dimensions (3D)\footnote{Indeed, the first-order term remains intact since it arises after the antisymmetrization of two indices. Therefore, we define it as $\cL^{\LL}_{m=1}\equiv(D-3)R$ to make this statement true for all $m$.}. Unfortunately, despite being very useful toy models for testing theoretical ideas beyond Einstein's theory,  this class of theories differs from GR only in higher dimensions ($D>4$). 

From the field theoretical perspective, one expects a better high energy behaviour in a lower dimension and, therefore, 3D gravity theories are candidates to give hints toward the solution of ``the real problem'' in 4D. One might think that this line of research is not promising since, in 3D, GR has no dynamical physical degrees of freedom around the flat space \cite{Deser:1983tn} and the LL-type modifications are not possible. However, non-trivial dynamics can be generated at the expense of having field equations with higher derivatives of the metric tensor. Among works in this vein, New Massive Gravity (NMG), which is obtained by adding a particular combination of quadratic curvature terms to the Einstein-Hilbert action, provides a non-linear completion of the Fierz-Pauli theory and describes a unitary excitation of massive spin-2 gravitons around the constant curvature vacua \cite{Bergshoeff:2009hq,Bergshoeff:2009aq} although having four-derivative field equations. Apart from the unitarity of fluctuations, the same quadratic invariant was also derived  by demanding the existence of a holographic c-function \cite{Sinha:2010ai}. In the same work, the cubic and quartic invariants having the same property was also derived, and later generalized to curvature invariants of arbitrary order in \cite{Paulos:2010ke}.  The unitarity of these theories  around maximally symmetric vacua  was confirmed up to cubic order \cite{Gullu:2010vw} and to all orders on \cite{Paulos:2010ke}. A unitary, Born-Infeld (BI)-type modification of NMG (BINMG) which gives NMG at the quadratic order in a small curvature expansion was constructed \cite{Gullu:2010pc,Gullu:2010vw}. Additionally, at higher orders, the expansion replicates the invariants constructed from the c-theorem in \cite{Sinha:2010ai}. The underlying reason was explained in \cite{Alkac:2018whk}: a BI-type theory admitting a c-function\footnote{For BINMG, it was first proved in \cite{Gullu:2010st}} must necessarily lead to such invariants at each order in the curvature expansion. The crucial ingredient in the holographic construction of these theories is that they still yield second-derivative field equations for the domain wall metric that is used to realize the holographic renormalization group flow under certain assumptions for the matter fields. In what follows, we will refer to these 3D theories as Sinha-Paulos (SP) theories and the corresponding Lagrangians,  denoted by $\cL^{\SP}$, as SP invariants, after their discoverers in the holographic context.

Having second-order field equations for any metric, LL theories naturally admit a holographic c-function \cite{Liu:2010xc}. Therefore, we have two sets of theories with the following properties: i) Both LL and SP theories admit a holographic c-theorem. ii) Around maximally symmetric vacua, LL theories describe unitary massless spin-2 excitations in $D>4$, and SP theories  describe unitary massive spin-2 excitations. 

The aim of this paper is to show the connection between the two theories at the Lagrangian level, which is rather unexpected, since the LL invariants vanish identically in 3D. This connection is achieved by taking a \textit{holographic limit} (h-lim) of LL invariants defined as
\begin{equation}
\hlim_{D\rightarrow \,3}\, \cL^{\LL}_m\equiv\lim_{D\rightarrow \,3}\frac{1}{D-3}\, \cL^{\RLL}_m,\label{eq:hlim}
\end{equation}
where $\cL^{\RLL}$ are the reduced-LL (RLL) invariants,  the part of LL invariants which contributes nontrivially in the formulation of the holographic c-theorem. Without a detailed explanation of the definition, which we will give later by explicit examples, let us state our main result: the holographic limit of the  LL invariants give SP invariants at each order $m$ as\footnote{SP invariants beyond $m=5$ contain free parameters, and  the holographic limit yields a particular choice of the free parameters. This will be elaborated after the discussion of the leading order invariants.}
\begin{equation}
\hlim_{D\rightarrow \,3}\, \cL^{\LL}_m=\cL^{\SP}_m.
\end{equation}
The backbone of this limit is the holographic c-function, a positive and monotonic function coinciding with the trace anomaly coefficients of the even-dimensional boundary field theory at infinity, which is derived under the assumption that matter fields in the bulk theory obey the null-energy condition (NEC) \cite{Freedman:1999gp,Myers:2010xs}. A simple method to derive the constraint imposed by the NEC can be utilized as follows \cite{Li:2017txk}: One starts with the following domain wall ansatz
\begin{equation}
ds^2=e^{2A(r)} \eta_{\hat{a} \hat{b}}\dif x^{\hat{a}}\dif x^{\hat{b}} +e^{2B(r)}\dif r^2,\label{eq:domain}
\end{equation}
where $\eta_{\hat{a} \hat{b}}$ is the Minkowski metric and the hatted-Latin indices run from $0$ to $(D - 2)$. One then considers the minimal coupling of a free scalar field to gravity
\begin{equation}
S=\int \dif^{D} x \sqrt{-g}\left[\cL^\text{gr}-\frac{1}{2} (\partial \phi)^2\right].\label{eq:scalaraction}
\end{equation}
Assuming $\phi = \phi(r)$ and using the ansatz (\ref{eq:domain}) in the action (\ref{eq:scalaraction}) give an effective action for the functions $(A(r),B(r),\phi(r))$. After finding the Euler-Lagrange equations and fixing the gauge by $B(r)=0$, the NEC is realized from the resulting equations as the positivity of the radial kinetic energy of the scalar field as follows
\begin{equation}
\frac{1}{2 } \phi^{\prime 2}=F\left(A^{\prime}, A^{\prime \prime}, \ldots\right) \geq 0,\label{eq:NEQ}
\end{equation}
where F is a polynomial of the derivatives of the function $A(r)$. For the theories admitting the holographic c-theorem, the condition takes the following form
\begin{equation}
\sum_{n=0}^{D-1} a_{n} A^{\prime 2 n} A^{\prime \prime} \geq 0,\label{eq:ineq}
\end{equation}
with second-order derivatives at most. A close scrutiny shows that a monotonic function can be derived from the inequality (\ref{eq:ineq})  (see \cite{Sinha:2010ai,Paulos:2010ke,Alkac:2018whk,Gullu:2010st,Liu:2010xc,Freedman:1999gp,Myers:2010xs,Li:2017txk,Myers:2010tj} for explicit expressions). The crucial ingredient for us is that using the ansatz (\ref{eq:domain}) directly in the action (\ref{eq:scalaraction}) allows  to distinguish the part of the action that contributes to the NEC (\ref{eq:NEQ}).

Now, we are in a position to show how the holographic limit works. All the Lagrangians that we consider will be expressed in terms of the Ricci scalar and the trace of the powers of the traceless Ricci tensor $\tR_{a b}=R_{a b}-\frac{1}{D}g_{a b}R$, which is defined as $\cR_n=\tR^{a}_{j_1}\tR^{j_1}_{j_2}\ldots\tR^{j_{n-1}}_{a}$.
Note that, in 3D, there are no independent $\cR_n$ invariants beyond $n=3$ \cite{Paulos:2010ke}. To demonstrate the main aspects of the limit, we consider SP invariants of order\footnote{For $m=1$, the limit leads to the Einstein-Hilbert term trivially.} 2, 3 and 4 
\begin{align}
\cL^{\SP}_{m=2}=&\frac{R^2}{24}-\cR_{2},\label{eq:SP2}\\
\cL^{\SP}_{m=3}=&\frac{R^{3}}{144}-\frac{R\,\cR_{2}}{2}+4 \cR_{3},\label{eq:SP3}\\
\cL^{\SP}_{m=4}=&\frac{R^4}{144}-R^2\cR_{2}-12(\cR_{2})^2+16R\,\cR_{3},\label{eq:SP4}
\end{align}
The LL invariants of the same orders are given as
\begin{align}
\cL^{\LL}_{m=2}=&(D-3)\left[\frac{(D-2)R^2}{4D(D-1)}-\frac{\cR_2}{(D-2)}\right]+\text{Weyl Terms},\label{eq:LL2}\\
\cL^{\LL}_{m=3}=&\frac{(D-2)!}{(D-6)!}\left[\frac{R^{3}}{8D^2(D-1)^2}-\frac{3R \,\mathcal{R}_{2}}{2D(D-1)(D-2)^2}+\frac{2\mathcal{R}_{3}}{(D-2)^4}\right]+\text{Weyl Terms},\label{eq:LL3}\\
\cL^{\LL}_{m=4}=&\frac{(D-3)!}{(D-8)!}\left[\frac{(D-2)R^4}{16D^3(D-1)^3}-\frac{3R^2\cR_{2}}{2D^2(D-2)(D-1)^2}+\frac{4R\,\cR_{3}}{D(D-1)(D-2)^3}\right]\nonumber\\
-&\frac{3(D-4)!}{(D-8)!(D-2)^4}\left[2\cR_{4}-(\cR_2)^2\right]+\text{Weyl Terms},\label{eq:LL4}
\end{align}
where Weyl terms denote  curvature invariants involving the Weyl tensor. These invariants are fixed up to an overall scaling, i.e. only the relative coefficients have physical importance. The vanishing of the LL invariants in $D=3$ is also apparent in this form since the Weyl tensor is zero for 3D metrics, the middle term in (\ref{eq:LL4}) vanishes in 3D due to the Schouten identity \cite{Paulos:2010ke} and other terms carry an explicit $(D-3)$ factor.

The next step is to construct the RLL invariants by identifying and then removing the parts of the LL invariants which do not play a role in the holographic c-theorem. The Weyl tensor vanishes for the domain wall ansatz since it is conformally flat. Therefore, the second and third order RLL invariants can be obtained by removing the Weyl terms as
\begin{align}
\cL^{\RLL}_{m=2}=&(D-3)\left[\frac{(D-2)R^2}{4D(D-1)}-\frac{\cR_2}{(D-2)}\right],\label{eq:RLL2}\\
\cL^{\RLL}_{m=3}=&\frac{(D-2)!}{(D-6)!}\left[\frac{R^{3}}{8D^2(D-1)^2}-\frac{3R \,\mathcal{R}_{2}}{2D(D-1)(D-2)^2}+\frac{2\mathcal{R}_{3}}{(D-2)^4}\right]
\label{eq:RLL3},
\end{align}
from which the holographic limit defined in (\ref{eq:hlim}) can be performed, leading to the second and third order SP invariants (\ref{eq:SP2}) and (\ref{eq:SP3}) as promised. The third order LL invariant (\ref{eq:LL3}) requires a special care. For the domain wall ansatz, the invariants $\cR_{4}$ and $(\cR_{2})^2$ are not independent and related by
\begin{equation}
D(D-1)\cR_{4}=(D^2-3D+3)(\cR_{2})^2.\label{eq:identity}
\end{equation}
Using this in (\ref{eq:LL4})  and then removing the Weyl terms yield
\begin{align}
\mathcal{L}^{\RLL}_{m=4}=&\frac{(D-3)!}{(D-8)!}\left[\frac{(D-2)R^4}{16D^3(D-1)^3}-\frac{3R^2\cR_{2}}{2D^2(D-2)(D-1)^2}+\frac{4R\,\cR_{3}-3(\cR_{2})^2}{D(D-1)(D-2)^3}\right]\label{eq:RLL4},
\end{align}
from which the holographic limit of the fourth-order LL invariant $\cL^{\LL}_{m=4}$ (\ref{eq:LL4}) can be taken, leading to the fourth order SP invariant (\ref{eq:SP4}).

The higher-order LL invariants are also expected to involve similar identities that make this limit possible. Here, the vitally important point is that the identity (\ref{eq:identity}) holds for the domain wall ansatz (\ref{eq:domain}), but not for all conformally flat metrics since it does not transform homogenously under conformal transformations.  Therefore, the holographic c-theorem must be the basis of this limiting procedure rather than the conformal flatness of (\ref{eq:domain}).

This prescription can be applied to theories more general than LL as long as they support a holographic c-function. One example is the Quasi-topological Gravity (QTG) \cite{Oliva:2010eb,Myers:2010ru,Myers:2010xs,Myers:2010tj}, which is a higher derivative theory constructed out of cubic curvature invariants. It is easy to show that the cubic part of it differs from the 3rd-order LL invariant $\cL^{\LL}_{m=3}$ only up to Weyl terms \cite{Oliva:2010eb}, and therefore it has the same holographic limit, which is just the third-order SP invariant $\cL^{\SP}_{m=3}$.

One important observation in 3D is that the SP invariants beyond $m=5$ are not uniquely determined because the constraints arising from the c-theorem are not enough to fix all the coefficients in the Lagrangian \cite{Paulos:2010ke}. Therefore, one might wonder what this implies in higher dimensions. If one constructs the most general Lagrangian $\cL^\MG_{(m>5)}$ admitting a c-function in $D$-dimensions ($D\geq4$), there will again appear some free parameters since the possibilities for forming a scalar out of curvature tensors increase with increasing $m$. When the holographic limit is taken, one should obtain the corresponding SP invariants with free parameters as
\begin{equation}
\hlim_{D\rightarrow \,3} \cL^\MG_{(m>5)}=\cL^{\SP}_{(m>5)}.
\end{equation}
The LL invariants with order $m>5$ form a subclass of these theories, where all the coefficients are fixed such that one has 2nd-order field equations for \textit{all} metrics. Therefore, their holographic limit should be as follows
\begin{equation}
\hlim_{D\rightarrow \,3} \cL^\LL_{(m>5)}={\bar{\cL}}^{\SP}_{(m>5)},
\end{equation}
where ${\bar{\cL}}^{\SP}_{(m>5)}$ denote a version of the SP invariants with fixed coefficients satisfying the constraints from the c-theorem. While starting from the most general Lagrangian yield the SP invariants in their full forms, taking the holographic limit of LL invariants leads to a sub-class of the SP invariants still admitting a c-function. The connection between 3D and higher-dimensional Lagrangians is still made by the holographic c-theorem but beyond $m=5$, the freedom due to the increasing number of possible curvature terms should be taken into account. A demonstration of this would be very useful. However, obtaining an explicit expression of $\cL^\LL_{(m>5)}$, which is necessary for performing the limit, amounts to evaluating the determinant of a $2m \times 2m$ matrix whose elements are Kronecker deltas. We leave this complicated task for future since the essence of the holographic limit is apparent at orders $m=2,\,3,\,4$.

Although similar in spirit, the holographic limit should not be confused with the recent work \cite{Glavan:2019inb}, where the authors claimed to obtain a novel Einstein-Gauss-Bonnet theory in 4D. The Gauss-Bonnet (GB) invariant, which is the lowest order LL invariant ($m=2$) after GR ($m=1$), is a boundary term in its critical dimension $D=4$ and therefore does not contribute the field equations. The observation in \cite{Glavan:2019inb} is that since the field equations come with a factor of $D-4$, for certain classes of spacetimes,  scaling the GB coupling by a factor of $\frac{1}{D-4}$ and then taking the limit $D\rightarrow4$ yield a non-trivial contribution and thereby new solutions. They considered constant curvature spacetimes,  the cosmological FLRW spacetimes, the static spherically symmetric black hole, and also the linearized fluctuations around maximally symmetric vacua. Although, these are physically relevant spacetimes which, in principle, lead to new observable predictions, a claim for a new theory requires the existence of the limit for \textit{any} solution of the theory in D-dimensions, which is unfortunately not the case \cite{Gurses:2020ofy,Ai:2020peo}. Indeed, a well-defined limit has been obtained in \cite{Fernandes:2020nbq,Lu:2020iav,Kobayashi:2020wqy,Hennigar:2020lsl}, giving a scalar-tensor theory, not a novel pure gravity theory in 4D, and also in 3D \cite{Hennigar:2020fkv}. This conclusion is also supported by the study of scattering amplitudes \cite{Bonifacio:2020vbk} (see \cite{Mahapatra:2020rds,Arrechea:2020evj} for other criticisms).

It is clear that defining a holographic limit to 4D as
\begin{equation}
\hlim_{D\rightarrow \,4}\, \cL^{\LL}_m\equiv\lim_{D\rightarrow \,4}\frac{1}{D-4}\, \cL^{\RLL}_m,\label{eq:hlim4}
\end{equation}
and applying it to the second order RLL invariant (\ref{eq:RLL2}) does not lead to a sensible result (see eqn-(17) of \cite{Liu:2010xc} for the explicit form of the function $F$ in (\ref{eq:NEQ}) and (\ref{eq:ineq})) while applying it to higher order invariants (\ref{eq:RLL3}) and (\ref{eq:RLL4}) yield Lagrangians in 4D which admit a c-function.  This shows explicitly that our work is fundamentally different than \cite{Glavan:2019inb}.
\begin{figure*}[!htb]
	\begin{center}
		\scalebox{0.46}{\includegraphics{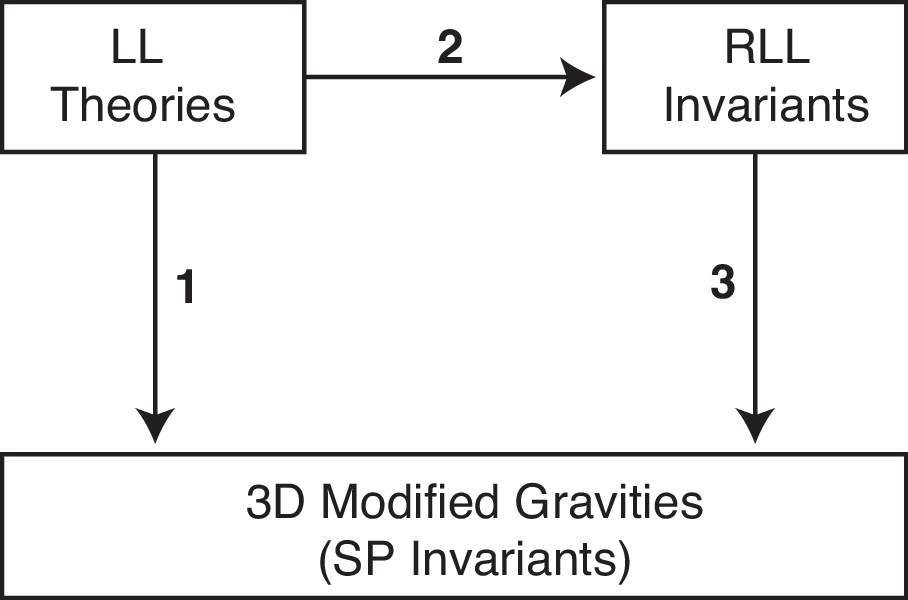}}
	\end{center}
	\caption{The prescription for taking the holographic limit: Map-1 is equivalent to Map-2 followed by Map-3}
	\label{fig} 
\end{figure*}

In conclusion, we have established a limiting procedure relating the higher and lower dimensional modified gravity theories at the Lagrangian level. Our results can be summarized as follows (see Figure-\ref{fig}):
\begin{enumerate}
	\item The holographic limit (Map-1) from LL theories to 3D modified gravity theories  is defined through the RLL invariants, which are the part of higher dimensional Lagrangians giving non-trivial contributions in the construction of the holographic c-function.
	\item Once the relevant RLL invariant is constructed (Map-2), SP invariants leading to 3D modified gravity theories can be easily obtained by removing a $(D-3)$ factor and then setting $D=3$ (Map-3)
	\item We have confirmed the limit for a higher-dimensional, higher-derivative theory, QTG, which suggests that the limit builds a relation between gravity theories in higher and lower dimensional theories in a general setup.
\end{enumerate}

Note that the SP invariants has been recently obtained in \cite{Gabadadze:2020tvt} by dimensional reduction of the LL invariants after a certain regularization. The connection between this approach and ours remains to be explored.\\

\noindent {\bf Acknowledgments}	D. O. D.  was supported in part by the National Natural Science Foundation of China under Grant No. 11875136 and the Major Program of the National Natural Science Foundation of China under Grant No. 11690021. We thank the anonymous referee for useful comments that led us to clarify an important point regarding the limit of higher order LL invariants.

\end{document}